\documentclass[3p]{elsarticle}

\usepackage{amssymb}
\usepackage{amsmath}
\usepackage{lineno}
\usepackage{xcolor}
\journal{Applied Mathematics and Computation}

\begin{document}


\begin{frontmatter}

\title{Mercenary punishment in structured populations}

\author[as]{Hsuan-Wei Lee}
\author[as]{Colin Cleveland}
\author[mfa]{and Attila Szolnoki}
\address[as]{Institute of Sociology, Academia Sinica, Taiwan}
\address[mfa]{Institute of Technical Physics and Materials Science, Centre for Energy Research, P.O. Box 49, H-1525 Budapest, Hungary}

\begin{abstract}
Punishing those who refuse to participate in common efforts is a known and intensively studied way to maintain cooperation among self-interested agents. But this act is costly, hence punishers who are generally also engaged in the original joint venture, become vulnerable, which jeopardizes the effectiveness of this incentive. As an alternative, we may hire special players, whose only duty is to watch the population and punish defectors. Such a policelike or mercenary punishment can be maintained by a tax-based fund. If this tax is negligible, a cyclic dominance may emerge among different strategies. When this tax is relevant then this solution disappears. In the latter case, the fine level becomes a significant factor that determines whether punisher players coexist with cooperators or alternatively with defectors. The maximal average outcome can be reached at an intermediate cost value of punishment. Our observations highlight that we should take special care when such kind of punishment and accompanying tax are introduced to reach a collective goal. 
\end{abstract}


\end{frontmatter}

\section{Introduction}
\label{intro}

The lack of cooperation can be identified as the key problem in almost every big crisis of humanity \cite{hardin_g_s68,pennisi_s05}. Examples can be given starting from global warming to overexploitation of common resources, but even health issues, like fighting against a pandemic could also be mentioned here \cite{nordhaus_pnas10,couto_jtb20,szolnoki_epl16,chica_srep21,wang_x_epl20,lee_hw_pa21,nagatani_pa19e,he_n_amc19}. Accordingly, it has paramount importance to identify those conditions and mechanisms which may inspire altruistic acts among self-interested agents. For a higher individual benefit it is tempting to betray our partner, therefore to aid cooperators or to lower the income of defectors could be a solution to resolve the mentioned social dilemmas \cite{fu_mj_pa21,szolnoki_epl10,cheng_f_amc20}. The latter can be done via a punishment where the personal income of those who do not want to contribute to the common pool may be reduced \cite{cong_r_srep17,gao_sp_pre20,yang_hx_epl20,lv_amc22,flores_jtb21,liu_jz_csf18}. For completeness, we note that there are alternative ways of punishment. For example defectors can be excluded directly from the benefit of joint ventures \cite{liu_lj_srep17,zhao_q_ijmpc20,xu_xg_pre17,szolnoki_pre17,quan_j_c19}.
Theoretically, we distinguish between peer and pool punishments \cite{chen_xj_pre15,szolnoki_pre11b,ohdaira_srep17,chen_xj_pcb18,quan_j_csf21}. In the former case, a cooperator penalizes a defector partner by facing the conflict and its circumstances personally. In the alternative case cooperators who bear the extra cost of punishment contribute to a centralized authority who punishes defectors in an institutionalized way. Independently of whether we apply peer or pool punishment, the sensitive part is the extra cost of the incentive, which establishes a higher level of the original social dilemma \cite{fehr_n04,helbing_ploscb10,hilbe_pnas14}. In particular, some fellows, who still contribute to the traditional pool, may deny to maintaining punishment, but still enjoys the positive consequences of the penalty. They can be considered as second-order free-riders and their success, by collecting higher income, will restore the original situation when pure cooperators and defectors compete again.

Unfortunately, this weakness cannot be avoided because monitoring others' activities requires extra work that should be considered via an additional cost in a model setup. We may respect such effort and in turn from those who bear the special cost of punishment we do not necessarily expect to contribute to the original joint venture. But we still allow them to enjoy its positive consequences. In other words, we pay them to guard the population and penalize those who just want to utilize the common goods. To distinguish this way of punishment from the previously mentioned traditional peer and pool punishments, we call the paid punishing actors mercenary players and their activity as mercenary punishment. Evidently, to keep these fellows requires an additional pool maintained by everyone independently of their choices of strategies. To justify the last condition, one might say that these ``armed" fellows have the power to collect the tax from everyone \cite{xu_l_csf18}. The core of this idea was previously introduced by Wang, Liu, and Chen, who explored its possible consequences in the framework of replicator dynamics \cite{wang_sx_pla21}. Their main finding was that the system evolves toward a full defection state in the absence of the mentioned tax, but the usage of tax-based punishment can drive the system to a state where cooperation survives.

The goal of our present work is to study their model in structured populations. This task is inspired by the fact that in structured populations, where players have limited access to other partners, the evolutionary outcome could be significantly different from the solution which is based on an assumption that the agents are mixed and their interactions are random \cite{nowak_n92b,santos_prl05,szabo_pr07,roca_plr09,amaral_pre20,perc_pr17,richter_bs19,liu_rr_amc19,lutz_pre21}. This was a fruitful and highly motivating revelation of last two-decade research and several studies confirmed that we should take special care when we try to extrapolate our observations obtained from a well-mixed population. 

In the following, we not just leave this comfortable condition that makes analytical calculations feasible, but we also explore the parameter space systematically to provide a comprehensive view about the possible consequences of tax-based or mercenary-type punishment. Our results confirm our previous expectations because in a more realistic structured population the solutions are richer including series of phase transitions by varying only a single control parameter. Furthermore, our work also stresses that an adequately chosen fine level could be a decisive factor, in which direction the system evolves during the evolutionary process. 

The remainder of this paper is organized as follows. In the next section, we summarize the key elements of the model and briefly discuss its well-mixed behavior. Next we present our observations obtained in structured populations and give further details about the main mechanisms which determine the evolutions. Last we summarize the main conclusions and discuss
their potential implications.

\section{Feeding a population from competing pools}
\label{def}

We consider an $L \times L$ square lattice with periodic boundary conditions where players are distributed on the nodes and interact with their neighbors. Players may choose from three strategies and can be a cooperator ($C$), a defector ($D$), or a punisher ($P$). According to the traditional public goods game setup, neighbors form $N=5$-member groups, a focal player and its four nearest neighbors, and decide simultaneously whether to contribute to a common pool or not \cite{brandt_prsb03,szolnoki_pre09c,perc_jrsi13}. While a cooperator provides a $c=1$ amount to the mentioned pool, neither a $D$ nor a $P$ player does. The collected contributions are multiplied by a synergy factor $r$ and the enhanced amount is distributed among the group members equally independently of their inputs. This is a classical social dilemma situation where it is better not to contribute but only enjoy the positive consequence of others' efforts. What distinguishes $D$ and $P$ players is the fact that the latter is monitoring the group and punishes a defector player whose payoff is reduced by a $\beta$ fine. As we already argued, such an activity, including the watching and the proper punishment, is costly, which is considered via a $G_P$ amount reduced from the punisher's income. To compensate for this cost all members of the population pay a $T$ tax to an alternative pool. One may claim that according to a defector's attitude, they are reluctant to pay such a tax, but we assume that punishers have the ability to collect it from everyone. This makes the main difference between the original and punishment's pools.

Summing up, the payoff values of different strategies originated from a game are the following 
\begin{eqnarray}
\label{pC}
\Pi_C &=& \frac{rc(N_c+1)}{N} - c - T\,,\\
\Pi_D &=& \frac{rcN_C}{N}-\beta N_P - T\,,\\
\Pi_P &=& \frac{rcN_C}{N} - G_P + \frac{NT}{N_P+1} - T\,,
\label{pP}
\end{eqnarray}
where $N_C$ and $N_P$ denote the number of other cooperators and punishers in the group respectively \cite{wang_sx_pla21}. We should stress that on a square lattice topology a player is involved in $N=5$ groups, in a one where it is a focal player and in four other games organized by the neighbors. Therefore the total income of every player is accumulated from the mentioned games.

As we noted, the payoff definitions described by Eqs.~(\ref{pC}-\ref{pP}) establish two independent pools. A traditional one which is supported by only $C$, but enjoyed by everybody, and a tax-based one which is supported by everyone, but enjoyed only by $P$. In turn, $P$ has a fix extra $G_P$ cost, signaling its permanent alertness to watch the population and punish free-riders. Evidently, to apply an incentive, like punishment, makes only sense if cooperators cannot survive otherwise. For the applied square lattice topology cooperators can coexist with defectors if the synergy factor $r$ exceeds a critical $r_c \approxeq 3.74$ value \cite{szolnoki_pre09b}, which is a straightforward consequence of network reciprocity. Therefore, in agreement with a previous work of Ref.\cite{wang_sx_pla21} we fix the synergy factor at $r=3$, which would yield a pure $D$ state in the classic model. In our model the remaining free parameters are $T$, $\beta$, and $G_P$. For simplicity, but not losing generality, we consider $T=0$, and $T=0.2$ cases, where the former is the limit of light tax, while the latter value represents the cases when the obligatory tax is significant. In the following we vary the values of the fine $\beta$ and the punishment cost $G_P$ systematically and explore the possible evolutionary outcomes.

The microscopic dynamics which determines the strategy update is the widely accepted pair-comparison based imitation \cite{szabo_pre98}. More precisely, a player $x$ enforces its strategy $s_x$ onto player $y$, who is having strategy $s_y$, with a probability which depends on their payoff difference 
\begin{equation}
\Gamma(s_x \to s_y)=1/\{1+\exp[(\Pi_{s_y}-\Pi_{s_x}) /K]\,\,.
\end{equation}
Here $K$ is a noise parameter, which allows an irrational strategy choice with a small probability. As in general, the presence of noise describes that we are not perfect and makes it possible for the system not to be trapped in an artificial state which would provide just a local optimum for the evolution. For the chance of comparison with previous model studies here we apply $K=0.5$ noise level. However, we declare that qualitative similar system behavior can be obtained if we apply other value of noise level. In particular, the qualitative topology of the presented phase diagrams remain unchanged, the only difference is the accurate positions of the phase transition points. The decisive factor, as we argue later, is the value of additional tax $T$. 

In our Monte Carlo simulations, we have applied different system sizes ranging from $L=200$ to $L=800$ where the typical relaxation time was between $10^5$ and $10^6$ MC steps. During a step on average every player has a chance to update its strategy. 

Before presenting our observations we briefly summarize the results obtained on the basis of replication dynamics by assuming a well-mixed population \cite{wang_sx_pla21}. As Wang {\it et al.} reported, in the absence of an additional tax, the system evolves into a state where all players choose defection. When the obligatory tax is considered, a new stable solution emerges where punishers and cooperators coexist, hence tax-based pure punishment can promote the emergence of altruistic behavior. As we will see, a structured population offers a significantly richer behavior and allows us to make more accurate predictions on how to design a society for general well-being.

\section{Results}

By following our previously declared concept, we are interested only in the situations when the tragedy of the common situation is inevitable in the absence of an additional incentive, like punishment. This happens when the value of synergy factor is too low therefore even the impact of network reciprocity is insufficient. In these cases, the aggregation of cooperators is unable to provide a competitive payoff for altruistic players to fight against defectors. Therefore additional strategy or other mechanism is necessary to keep cooperation alive.

\subsection{Mercenary punishment in the low tax limit}

\begin{figure}[h!]
\centering
\includegraphics[width=7.5cm]{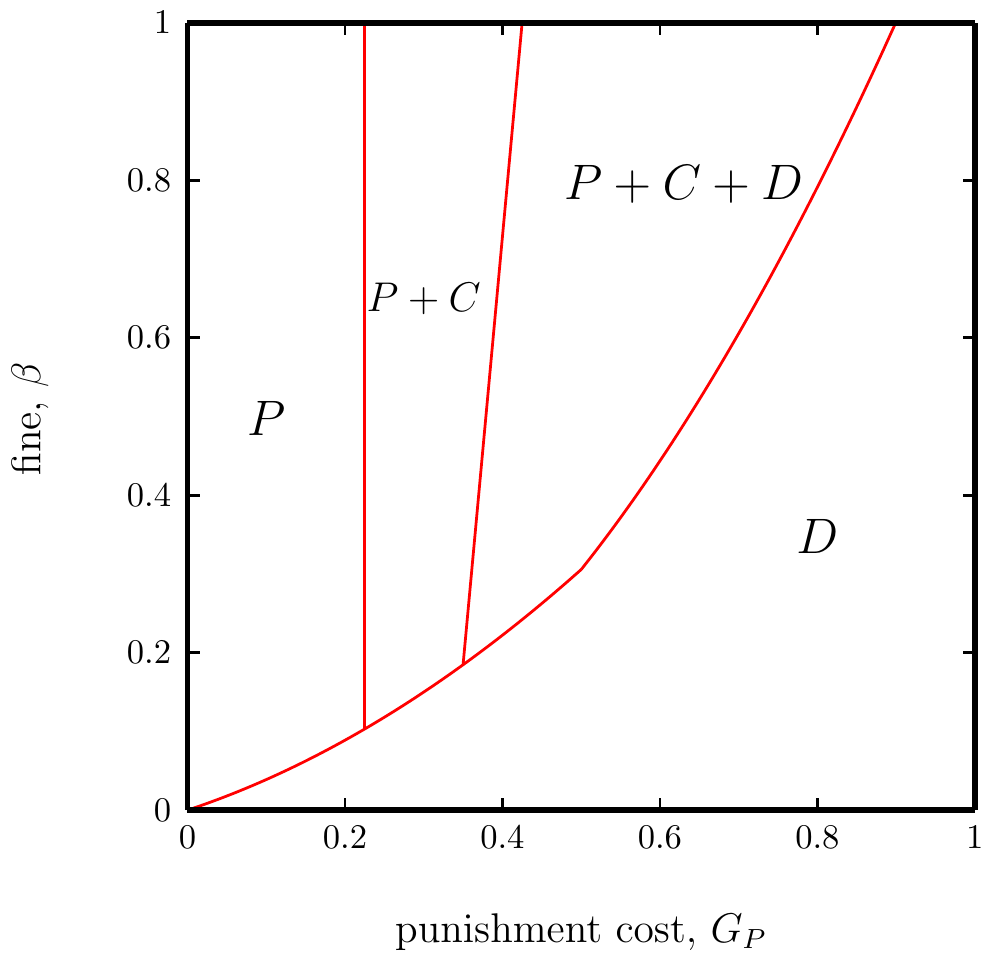}\\
\caption{Schematic phase diagram on the punishment cost -- penalty plane obtained at $r=3$, $T=0$. Despite the lack of tax, cooperators can coexist with punishers at an intermediate cost value if the fine is high enough. When we weaken punishers further by increasing the punishment cost then a cyclic dominance emerges, where $C \to P \to D \to C$ is the rank among competing strategies. By increasing $G_P$ further, this delicate balance is broken and defectors prevail.}\label{phd_T_0}
\end{figure}

First, we present the results of the proposed model obtained in the zero tax limit. In his case punishers have no right to impose an extra tax on the population. Still, they are supported by cooperators because they also enjoy the benefit of the common pool without contributing it. This is why we call this kind of punishment mercenary to distinguish it from previously introduced peer and pool punishment where the punishers also take share the original duty and also contribute to the common pool. One may claim that defectors have a similar advantage by avoiding the cost, but punishers have the skill to penalize them for free-riding. Therefore it could be essential how high cost should bear by punishers and how large fine is imposed by them. To explore all possibilities we scan the full parameter space in dependence of the mentioned parameters. Our results are summarized in Fig.~\ref{phd_T_0}, where we plot the schematic phase diagram. It means that we designate those solutions where the system evolves from a random initial state if the system size is large enough. Notably, the latter criterion could be a crucial factor because in some cases the final destination could be different from those shown in our figure due to too small system size. In the following we will give further details about this annoying effect.

Turning back to the phase diagram, the message is crystal clear. If the punishment cost is low enough and the applied penalty is large enough then defectors die out from the population. More precisely, at very low cost and high fine $P$ strategy prevails and forms a homogeneous system. But this success is questionable because in the absence of cooperators they miss the benefit of the original pool. Ironically, a larger punishment cost brings them salvation because in this case, the stable solution of $P$ and $C$ strategies emerges where the average income of players grows. We note that in a well-mixed population this solution was only reported for significant tax level \cite{wang_sx_pla21}. Interestingly enough, when we weaken the $P$ strategy further, by increasing the value of $G_P$ more, then a new kind of solution appears. In this case all three strategies coexist. As expected, however, if we assume drastically high $G_P$ then punishers become too weak and they die out. As a consequence, only $D$ and $C$ strategies remain whose fight eventually leads to the total victor of free-riders due to the low value of $r$. Summing up, to hire mercenaries whose only duty to punish defectors may work even in the absence of additional tax, but only if the punishment cost is bearable. 
 
\begin{figure}[h!]
\centering
\includegraphics[width=8.5cm]{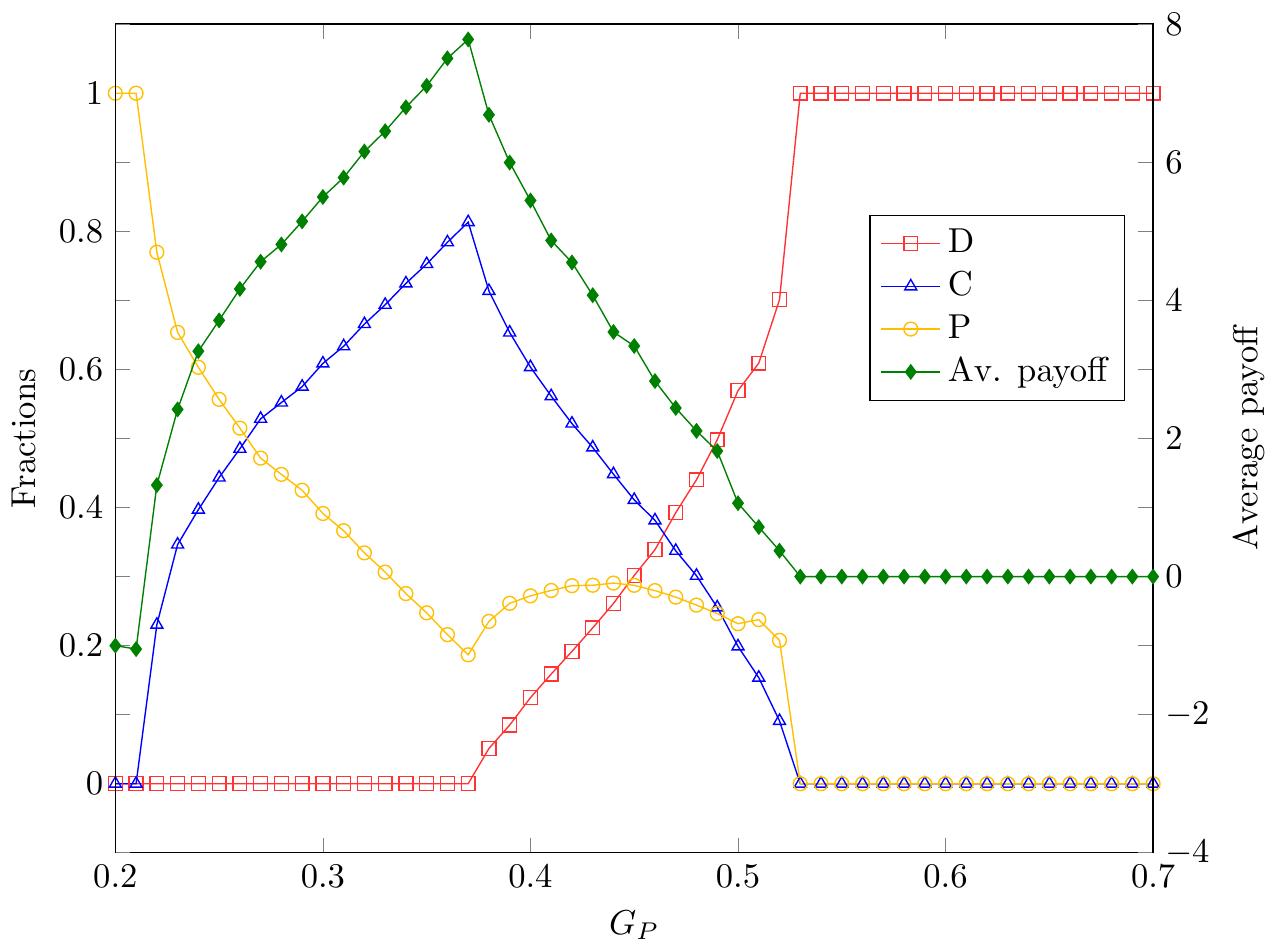}\\
\caption{Cross section of the phase diagram shown in
Fig.~\ref{phd_T_0}, as obtained for $\beta=0.4$ fine value. Depicted are stationary fractions of the three competing strategies as a function of punishment cost. As we increase $G_P$, the system enters to $P+C$ solution from full $P$ state, then $P+C+D$ state emerges, and finally it arrives to full $D$ state. We also plotted the average payoff values of the population, which signs that there is an optimal $G_P$ cost level of punishers which is necessary to reach the best collective income.}\label{cros_T_0}
\end{figure}

To get a deeper insight into the emerging solutions we next present a cross section of the phase diagram obtained at a fixed $\beta=0.4$ fine value. Fig.~\ref{cros_T_0} shows the stationary fractions of strategies in dependence of the punishment cost $G_P$. With a small cost value, punishers dominate. As we increase this cost, we enter the desired $P+C$ phase where the fractions of cooperators, hence the general well-being grows by increasing $G_P$. Above a threshold cost level, however, the situation changes and defectors appear again. This is because cooperators become too strong comparing to punishers, or put differently, punishers become weaker, hence cooperators can beat punishers. This fact establishes a cyclic dominance between the three strategies, because $P$ strategy is still stronger than $D$, while the latter strategy always dominates cooperators. In this way a cyclic $C$ beats $P$ beats $D$ beats $C$ relations appear,  which establishes the stable coexistence of competing strategies. This mechanism is well-known from rock-scissors-paper game \cite{szolnoki_jrsif14}, but more importantly, similar behavior can be observed in a significantly wider range of game-theoretical models where the relation of strategies is less obvious on the basis of the payoff definitions \cite{szolnoki_csf20b,szolnoki_epl20}. Figure~\ref{cros_T_0} warns us that this delicate balance vanishes above a critical cost level. If $G_P$ is too large then punishers cannot dominate defectors anymore, hence the latter strategy will beat both remaining partners. This consequence, however, is natural because if the punishment cost is too large then punishment cannot compensate for this negative effect and the whole institution makes no sense anymore.
 
In the following we demonstrate that to detect the proper transition point between the $P+C+D$ and the full $D$ phases is numerically demanding which is accompanied by heavy finite-size effects. It is because cyclic dominance among competing strategies results in a heavy oscillation in small system size. More precisely, if the characteristic length of the merging patterns is comparable to the linear system size of the square grid then we can observe oscillation in the portion of strategies as we monitor their values. 

\begin{figure}[h!]
\centering
\includegraphics[width=6.5cm]{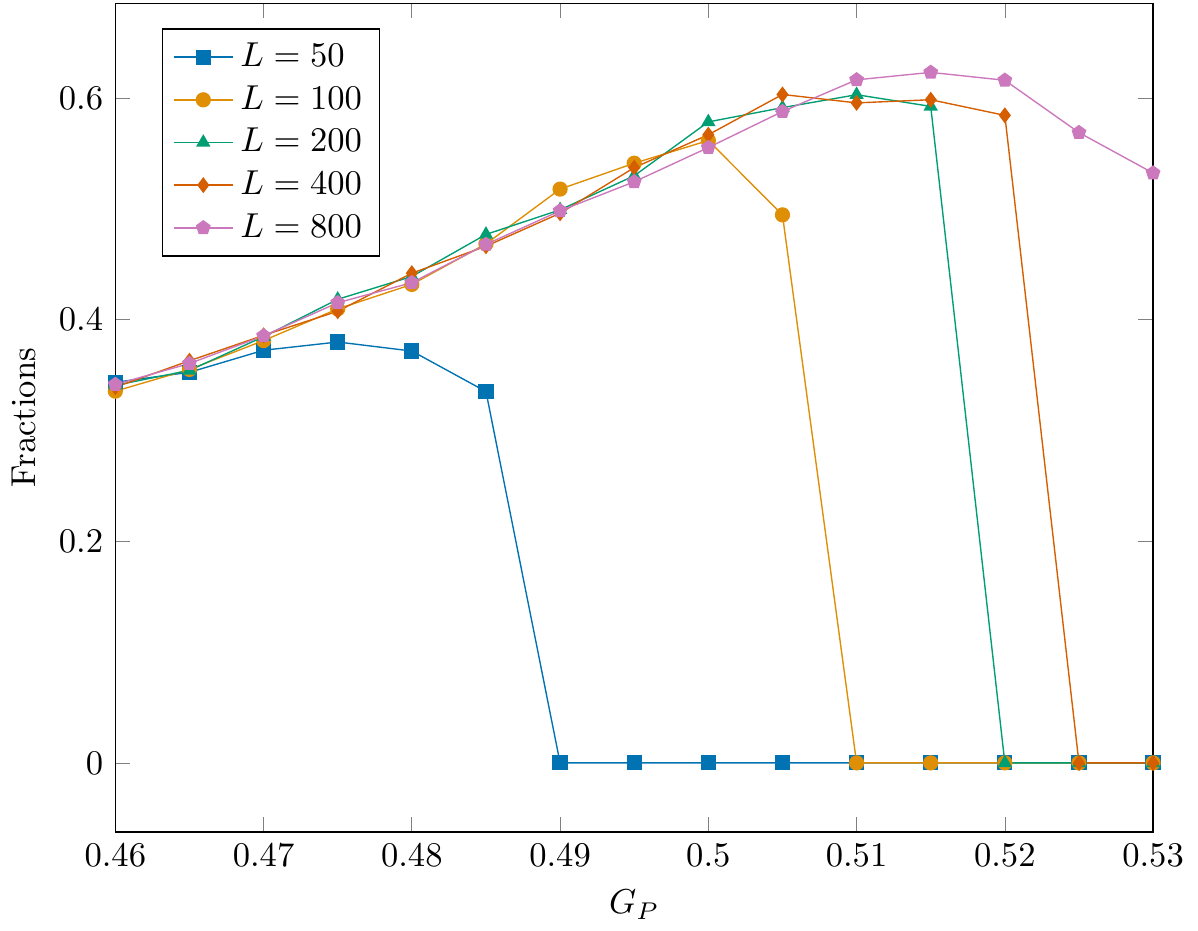}\\
\caption{Finite-size effect as we approach the border separating $P+C+D$ ad full $D$ phases. The stationary fraction of defectors is shown in dependence of punishment cost at $\beta=0.4$, $T=0$ for different sizes of the population (the liner size of the square grid is marked in the legend). Far from the transition point, even a small system is capable to give reliable predictions for the stationary state. But as we reach the critical region, the amplitude of the oscillation becomes so larger which can drive the system easily into a homogeneous destination. Evidently, the closer we are to the transition point the larger system size is necessary to observe the proper stationary state.}\label{fss}
\end{figure}

Importantly, such oscillation is completely missing in a spatial system if the size is large enough. As we vary the control parameter, which is $G_P$ in our present case, the relations of the strategies change, which results in more intensive invasion between the strategies. Accordingly, the amplitudes of the oscillation become larger which jeopardizes the coexistence of competitors. This effect emerges earlier for small system size, but can be avoided by using larger sizes. Interestingly, in the mentioned case a small-size simulation may result in a paradox outcome. For instance, if $C$ strategy dies out first then punishers can prevail and crowd out defectors despite the fact that we increased the value of $G_P$, hence we weakened the punisher strategy. This phenomenon is illustrated nicely in Fig.~\ref{fss} where we plot the stationary fractions of defectors in dependence of punishment cost. As we can see, for small system sizes we can easily obtain false destinations for the evolution process, but it can be fixed by using larger and larger system sizes.  

\begin{figure}[h!]
\centering
\includegraphics[width=6.5cm]{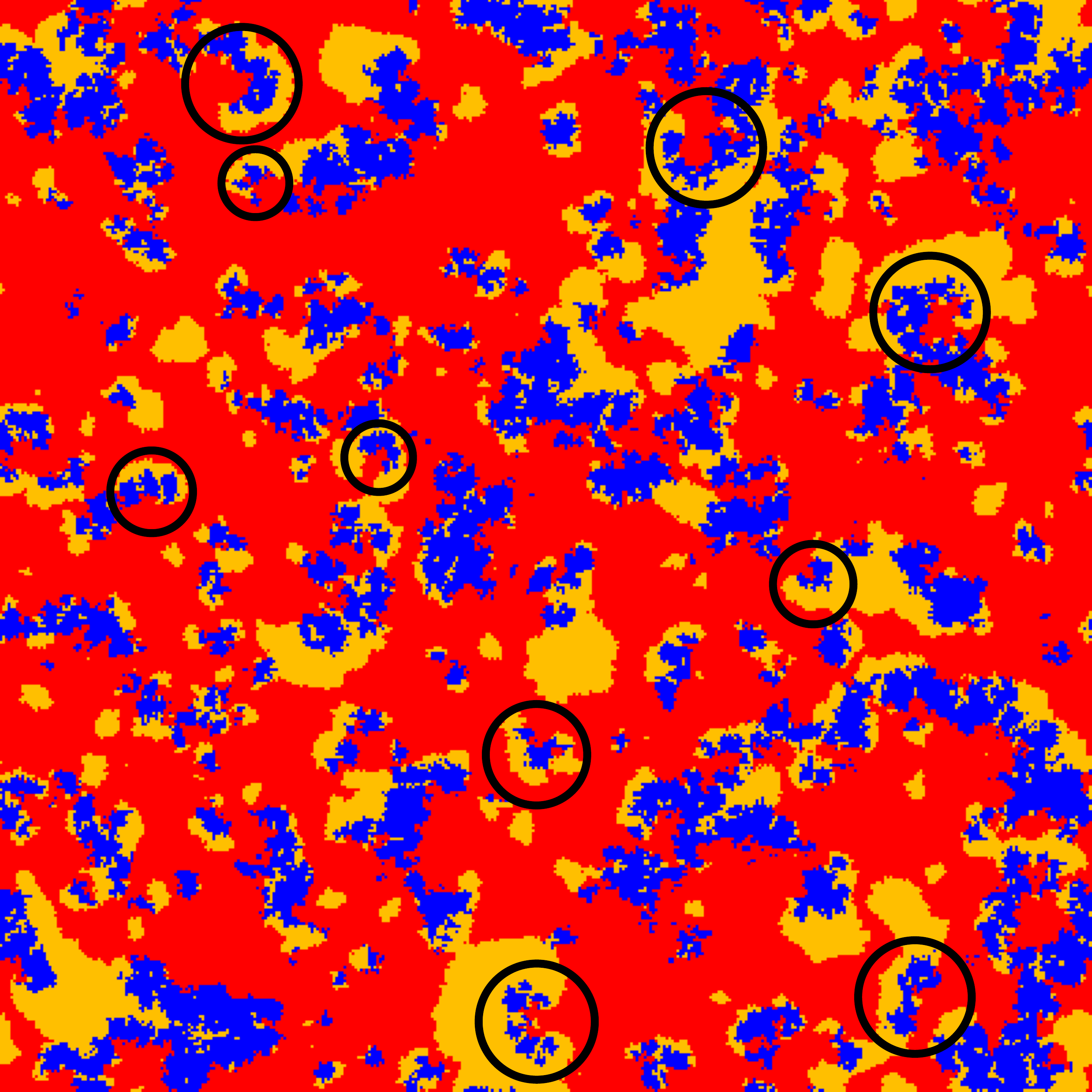}\\
\caption{Characteristic distribution of competing strategies in the $P+C+D$ phase. Blue, red, and yellow colors denote players with $C$, $D$, and $P$ strategy respectively. The rotating spiral patterns of invasion fronts between homogeneous domains signal clearly the cyclic dominance among three competing strategies. For clarity we marked them with black circles. The snapshot was taken at $\beta=0.4$, $T=0$, $G_P=0.5$ in the stationary state where linear size was $L=400$.}\label{pattern}
\end{figure}

To support our argument about the specific feature of the solution that characterizes the coexistence of all strategies in Fig.~\ref{pattern} we present a representative spatial distribution of strategies in the stationary state. The pattern shows domain walls separating homogeneous domains behave as propagating waves, and we can observe vortices in the centrum of rotating spirals \cite{szabo_pre99}. These are clear signs of a cyclic dominance in a spatial system, which were observed in rock-scissors-papers-type models \cite{jiang_ll_pla12,brown_pre19,avelino_epl21,serrao_epjb21,park_c18c}.

\subsection{Applying significant tax level}

\begin{figure}[h!]
\centering
\includegraphics[width=7.5cm]{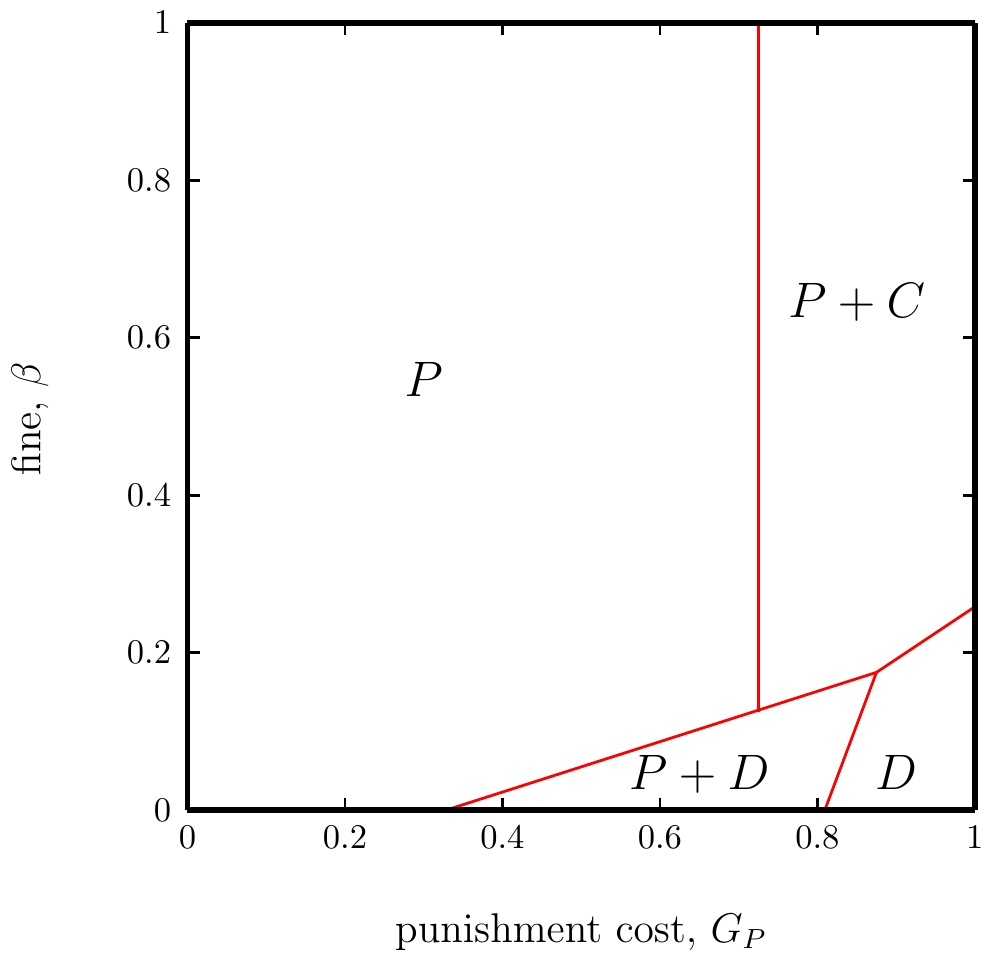}\\
\caption{Schematic phase diagram on the punishment cost -- penalty plane obtained at $r=3$, $T=0.2$. As expected, punishers benefit most from the introduction of an obligatory tax and they dominate the majority of parameter space. The previously observed $P+C$ solution still exists, but is shifted toward higher cost values. Interestingly, defectors may suffer the most from the tax, hence the cyclically dominant solutions vanish. The fine value, however, is a critical parameter, because at small fine values defectors can replace cooperators and coexist with punishers.}\label{phd_T_0_2}
\end{figure}

In this section we turn to the case when punishers are allowed to collect extra tax from the whole population for their punishing efforts and this tax level is substantial. Even if the value of $T$ is small, as Eqs.~(\ref{pC}-\ref{pP}) show, this extra contribution should be paid in every group interaction where the players are involved. The main observations are summarized in Fig.~\ref{phd_T_0_2} where we present a schematic phase diagram on the punishment cost and fine parameter plane. For proper comparison to the tax-free case, we here applied the same $r=3$ value which is too small to offer a chance for cooperators to survive in the absence of additional incentives. This diagram partly justifies our expectations. Namely, as we previously noted, this centralized tax is maintained by everyone, but only punishers benefit from it. Hence the parameter area where this strategy becomes dominant increases significantly. However, we should consider this system behavior as previously because we cannot be happy with the spreading of punishers in the absence of cooperators. While the previously reported $P+C$ phase still exists, but it has been shifted to larger $G_P$ cost values. We may interpret this shift as an obligatory tax is justified for high punishment cost values only. Indeed, it is also true that the parameter area where defectors can survive has shrunk remarkably. Interestingly, a new kind of stable solution emerges that was not observed previously. In particular, defectors can crowd out cooperators and coexist with punishers when the fine is low. This unfortunate scenario, when criminals coexist with the police, is a straightforward consequence of the fact that the imposed additional tax weakens cooperators more than defectors who have no additional duties.

\begin{figure}[h!]
\centering
\includegraphics[width=6.5cm]{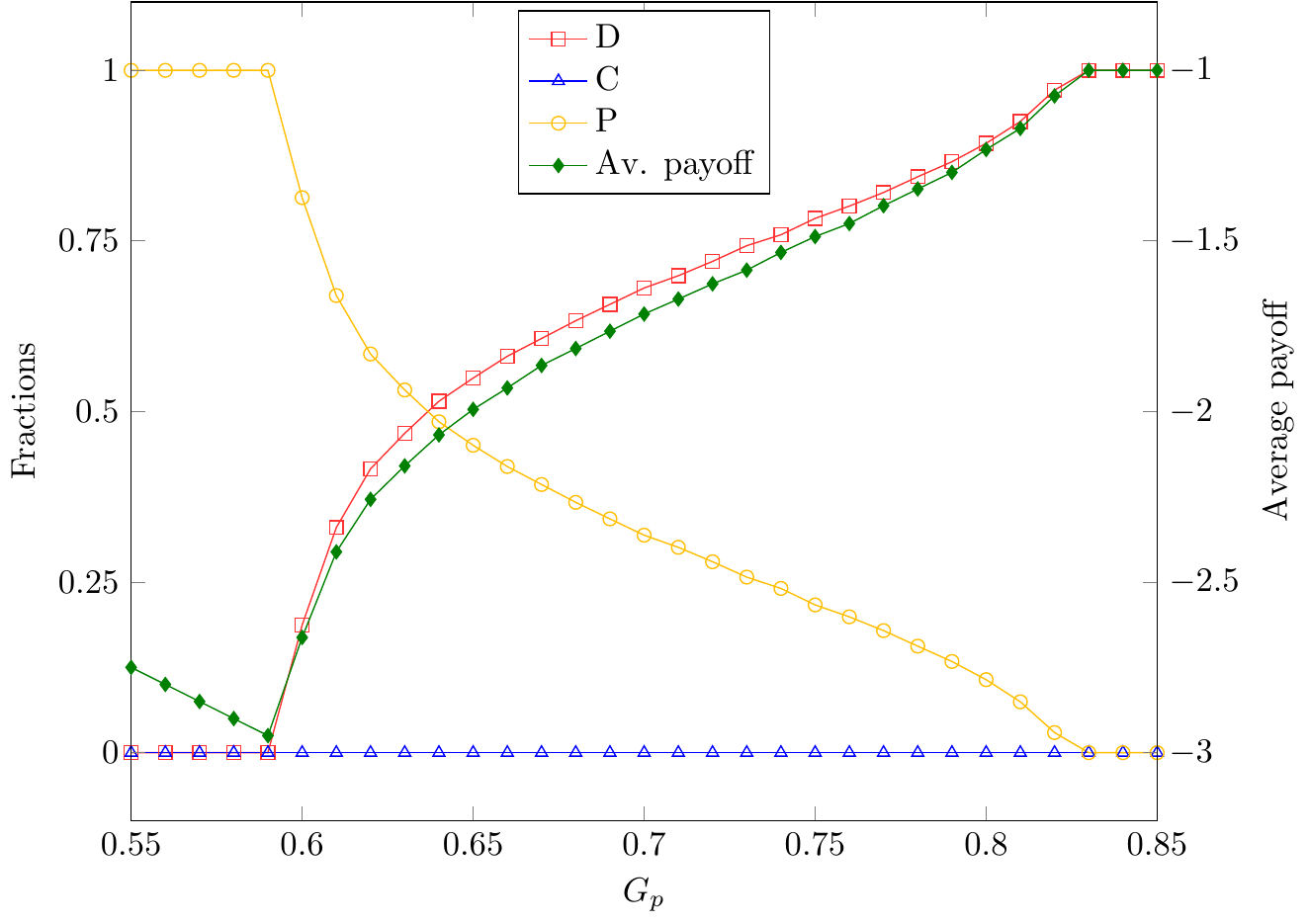}\includegraphics[width=6.5cm]{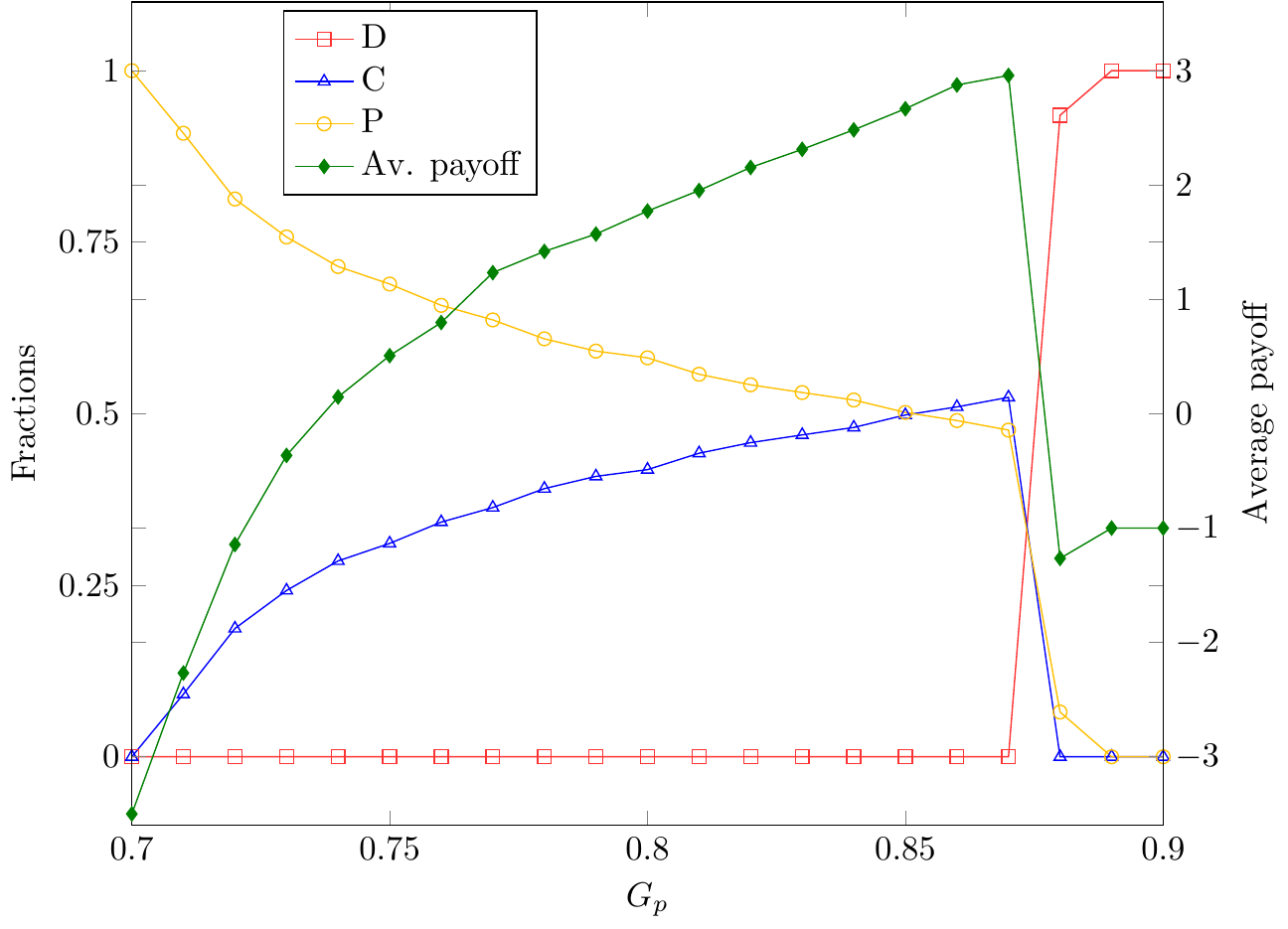}\\
\caption{Cross section of the phase diagram shown in
Fig.~\ref{phd_T_0_2}, as obtained for $\beta=0.1$ (left) and $\beta=0.2$ (right) fine values. Depicted are stationary fractions of the three competing strategies as a function of punishment cost. At low fine (left panel) the full $P$ and full $D$ states are separated by a phase where these strategies coexist. For larger fine values (shown in the right panel) the previous homogeneous phases are separated by $P+C$ and $P+D$ solutions. The average payoff values remain negative since there is no cooperators to contribute in the whole $G_p$ region we chose. In panel~(b) we also plotted the average payoff values of the population. Similar to the tax-free case, it signs that there is an optimal $G_P$ cost level of punishers when the highest collective income can be reached.}\label{cross_T_0_2}
\end{figure}

Similar to the previously discussed case we also present cross sections of the phase diagram. But in this case we show two lines for different fine values. Fig.~\ref{cross_T_0_2}(a) illustrates the situation when the fine value is low. Here, as we increase the punishment cost value, punishers will be gradually replaced by defectors and we can detect two continuous phase transitions when we leave the full $P$ state and arrives at the full $D$ destination. Fig.~\ref{cross_T_0_2}(b) shows a qualitatively different situation. If the fine is strong enough then the full $P$ solution is replaced by a $P+C$ coexistence as we increase the value of $G_P$. Of course, if the cost is too large then we get back again the full $D$ phase. But before it we can detect the $P+D$ phase again for a tiny range of punishment cost. Therefore we can identify three consecutive phase transitions until we reach the full $D$ phase again.

At first sight this very narrow $P+D$ phase seems to be invalid and one may suspect an insufficiently chosen relaxation time or other numerical inaccuracy. Therefore, in such cases, when we identify the stable solution which is valid in the large size limit, we should interpret the numerical outcomes with special care. More precisely, in the vicinity of a phase transition point, even conflicting results may be observed no matter we launch the simulations from a random initial state by using the same parameter values. This is again a clear sign of a finite-size effect. This problem can be handled even at system size which is numerically feasible if we check the stability of different solutions.

\begin{figure}[h!]
\centering
\includegraphics[width=6.5cm]{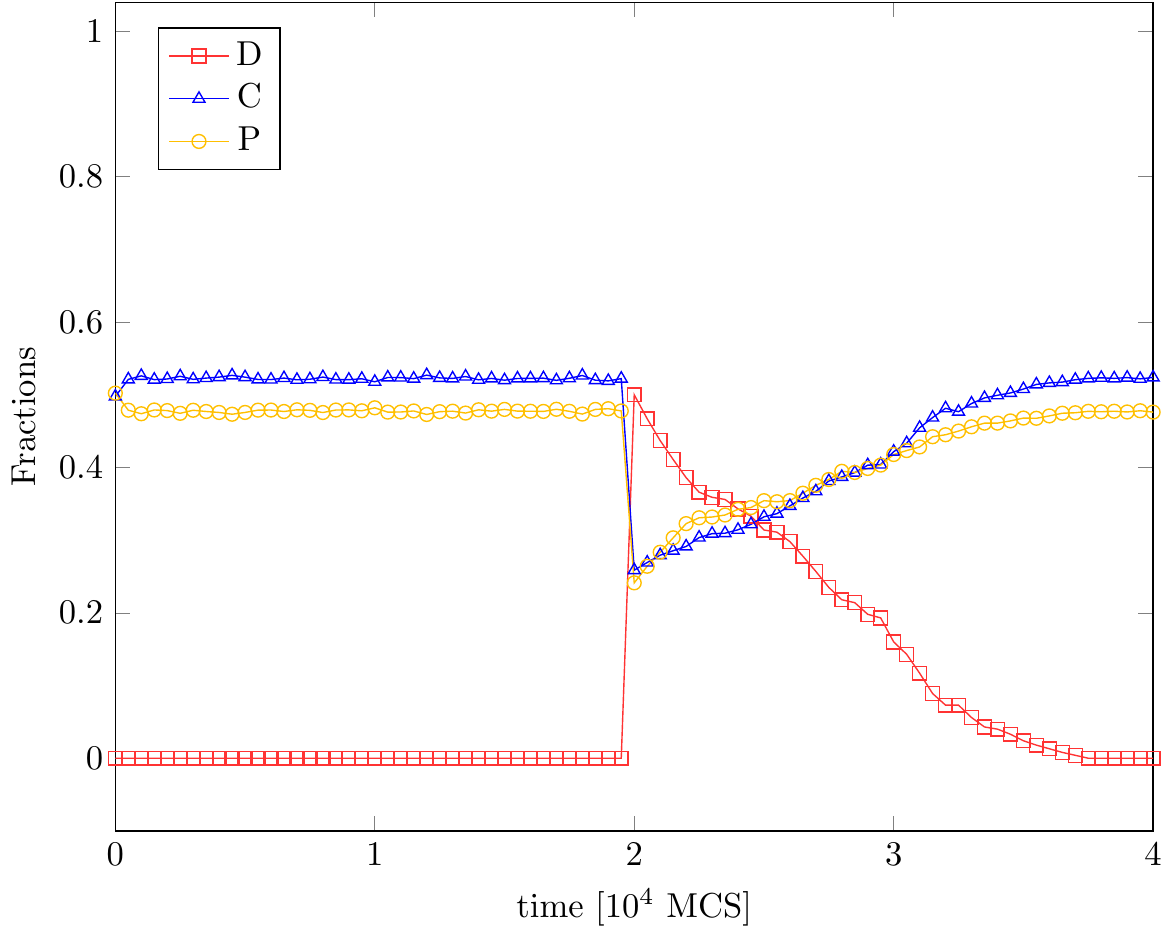}\includegraphics[width=6.5cm]{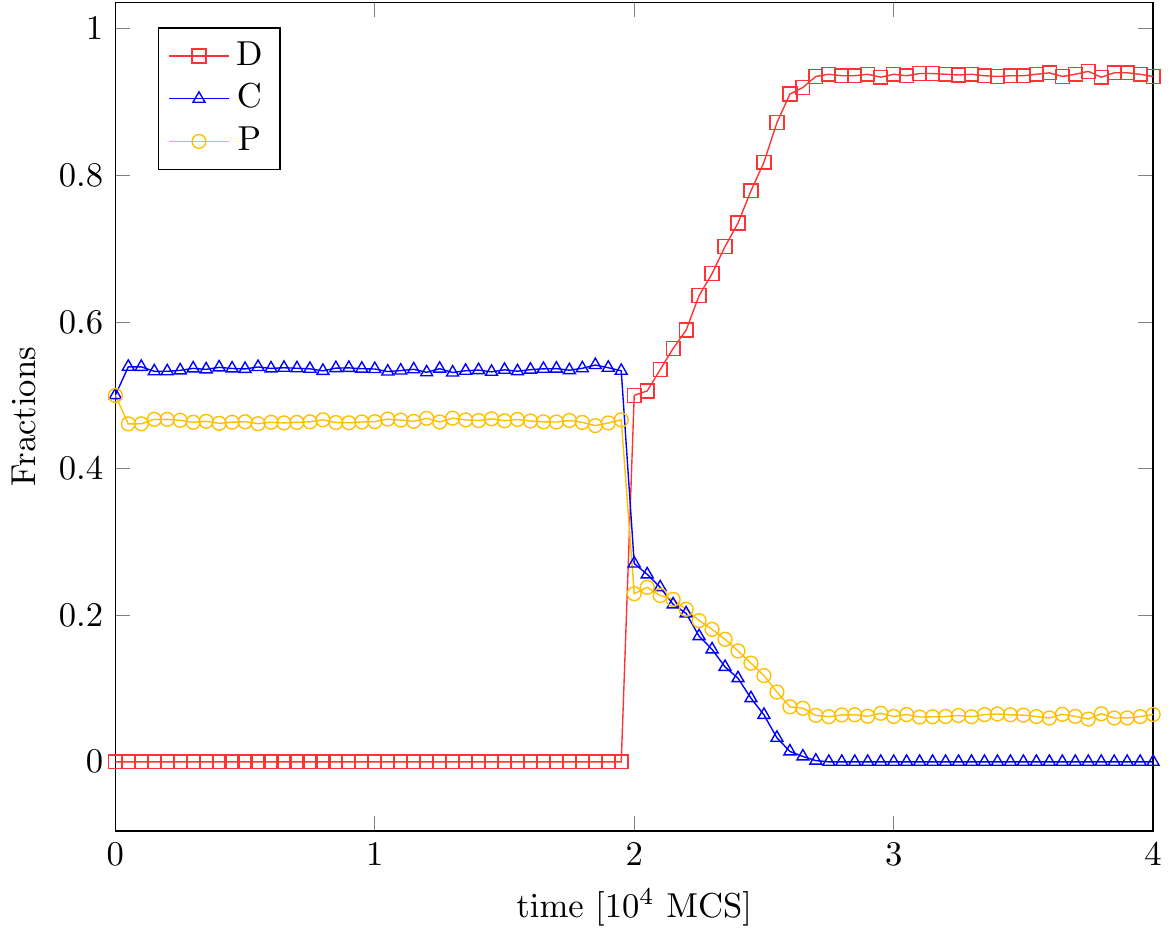}\\
\caption{Stability analysis of a solution at different parameter values. First we separate the system into two halves and on the left side, we allow $P$ and $C$ strategies to form a stable coexistence. Meanwhile, the other half of the available space is occupied by defectors only. After 20,000 MC steps, when the stable coexistence in the $P+C$ phase evolved, we allow the mentioned solutions to communicate with each other via strategy adoption. As a consequence, the $P+C$ solution will spread, shown in panel~(a), or a new $P+D$ solution emerges and dominate the whole system. The latter is shown in panel~(b). In both panels we present the time evolution of the strategy portions where the parameters are $r=3$, $\beta=0.2$, $L=400$, $G_P=0.87$ (left), $G_P=0.88$ (right panel).}\label{stab}
\end{figure}

This stability analysis is illustrated in Fig.~\ref{stab} where the same method is applied for two, albeit almost equal parameter values. In the beginning we separate the available space into two halves and fill the first half with $P$ and $C$ players only. The other half is occupied by defectors exclusively. Importantly, we do not allow the two halves of the population to communicate, but allow microscopic strategy invasion within the domains. As a result, a stable $P+C$ solution evolves in the first half where the fraction of cooperators exceeds the portion of punishers a bit. After, we open the borders and allow two domains to communicate via imitation steps. The only difference between panel~(a) and (b) is that we modified the cost value gently and increased it by $0.01$. This slight change of a parameter results in a dramatic change in the destination. While for smaller cost values the $P+C$ solution remains stable and finally dominates the whole population, for a bit higher cost a new solution emerges and defectors replace cooperators to form a stable coexistence. Interestingly, the portion of defectors is significantly larger than the fraction of cooperators in the $P+C$ solution. This behavior warns us that a small change in cost value may cause a dramatic change in the system behavior.

\section{Conclusion}

Does it pay to keep agents whose only duty is to watch the population and punish defectors? Shall we accept to pay an extra tax for their service? These are the questions which motivated us to study the possibility of mercenary punishment, as an alternative way of negative incentives to control defectors in a society. This can be done in a three-strategy model in which we extend the classic public goods game of cooperators and defectors by adding punishers. Like everyone else, the latter players also enjoy the benefit of the common good, but they do not contribute directly. Instead, they focus on watching the population to identify and punish defectors. Besides the traditional pool, an alternative tax-based pool may also be considered which serves to cover the punishers' expenses directly. Importantly, the model was already introduced by Wang, Liu, and Chen who studied it in a replicator dynamics approach. Our prime motivation by revisiting their model was to explore whether the assumption that the population is structured offers new, previously not observed behavior. Notably, this last question becomes a standard one in the last decade because several previous models underlined that the change from a randomly mixed population to a more realistic one where players have limited interactions with their neighbors can alter the system behavior qualitatively \cite{szabo_jtb12,fu_y_c21,broom_dga21}.

In agreement with our preliminary expectation, the assumption that players interact in a structured way has brought several new solutions which were not identified in the well-mixed case. First, we note that the desired defector-free solution can be reached even in the absence of additional tax no matter the synergy factor is too low which would dictate a full $D$ state in the classic public goods game. Secondly, we have identified a solution here in which competing strategies dominate each other cyclically. We note that this solution can also be observed for low-tax cases. This finding underlines that such a solution can emerge not only in traditional rock-scissors-papers type game, but can be observed in a wider scale of systems which can be characterized by the concept of evolutionary game theory \cite{garde_rsob20,palombi_epjb20,baker_jtb20,vukov_pre13,avelino_pre19b,cheng_f_pa19,esmaeili_pre18}. Thirdly, we point out the rotating spirals pattern resulting from cyclically dominant strategies can easily cause serious finite-size problems. The oscillations of the quantities could be too large that lattices with small sizes couldn't yield an accurate result. We demonstrate this potential danger in a critical region.

Furthermore, the comparison of phase diagrams obtained at different situations highlights that taxation, as an extra way of punisher's support can only be justified if the punishment cost is significant. Otherwise, the mercenary-type punishment, when punishers enjoy the traditional pool freely, already provides the necessary help for punishers to survive and control defectors. Indeed, in the lack of such tax defectors dominate a considerable area of the parameter plane even at high fine values if the punishment cost is relevant. From this viewpoint the introduction of an obligatory tax, in other words, if we allow punishers to collect an extra tax from everyone, is capable of shrinking the area of the full $D$ phase in the parameter space significantly. But even in this case, a substantial fine should be applied, otherwise the extra expense becomes more detrimental for cooperators who are displaced by defectors. We might say that despite good will, punishers may hurt those who hire them to give space for those who should be the target of the punishment. Therefore, as a conclusion, when we try to handle defectors and use mercenary punishment then we should estimate the possible cost of punishment carefully and allow the involved actors to apply an extra tax only if this cost is really significant. But even in this case, we should demand them to impose a deterrent fine level on defectors for the expected evolutionary outcome.

\vspace{0.5cm}

This research was partially supported by the Ministry of science and technology of the Republic of China (Taiwan), under grant No. 109-2410-H-001-006-MY3.

\bibliographystyle{elsarticle-num-names}

\end{document}